\documentclass{aa}

\usepackage{graphicx}

\newcommand{\beq}{\begin{equation}}
\newcommand{\eeq}{\end{equation}} 
\newcommand{\eeqn}[1]{\label{#1}\end{equation}}
\newcommand{\BV}{Brunt-V\"ais\"al\"a\ }
\def\Ln{\mathop{\hbox{ln}}\nolimits}
\def\Div{\mathop{\hbox{div}}\nolimits}
\newcommand{\vu}{\vec{u}}
\newcommand{\lp}{ \left(}
\newcommand{\rp}{ \right)}
\newcommand{\na}{ \vec{\nabla} }
\newcommand{\vg}{\vec{g}} 
\newcommand{\disp}[1]{\displaystyle #1}
\newcommand{\dz}[1]{\frac{\partial  #1}{\partial z}}
\newcommand{\ltex}[1]{\quad \hbox{#1} \quad}
\newcommand{\eq}[1]{(\ref{#1})}
\newcommand{\dnz}[1]{\frac{d  #1}{dz}}
\newcommand{\greq}{\begin{equation}\left\{ \begin{array}{l}}
\newcommand{\egreq}{\end{array}\right. \end{equation}}
\newcommand{\vxi}{\vec{\xi}}
\newcommand{\ez}{\vec{e}_z}
\newcommand{\egreqn}[1]{\end{array}\right. \label{#1}\end{equation}}
\newcommand{\vi}{\vec{v}}
\newcommand{\vpsi}{\vec{\psi}}
\newcommand{\lc}{ \left[}
\newcommand{\rc}{ \right]}
\newcommand{\noi}{ \noindent }
\newcommand{\infapp}{\raisebox{-.7ex}{$\stackrel{<}{\sim}$}}
\newcommand{\ddnz}[1]{\frac{d^2  #1}{dz^2}}
\renewcommand{\descriptionlabel}[1]%
{\hspace{\labelsep}\textsf{#1}}

\begin{document}

\title{Identification of gravity waves in hydrodynamical simulations}

\author{Boris Dintrans\inst{1} \and Axel Brandenburg\inst{2}}

\authorrunning{Boris Dintrans \& Axel Brandenburg}

\offprints{dintrans@ast.obs-mip.fr}

\institute{Observatoire Midi-Pyr\'en\'ees, CNRS UMR5572, 14 avenue
Edouard Belin, F-31400 Toulouse, France \and NORDITA,
Blegdamsvej 17, DK-2100 Copenhagen \O, Denmark}

\date{\today}

\abstract{
 The excitation of internal gravity waves by an entropy bubble
 oscillating in an isothermal atmosphere is investigated using direct 
 two-dimensional
 numerical simulations. The oscillation field is measured by a
 projection of the simulated velocity field onto
 the anelastic solutions of the linear eigenvalue problem for the
 perturbations. This
 facilitates a quantitative study of both the spectrum and the amplitudes
 of excited $g$-modes.
\keywords{hydrodynamics - waves - methods: numerical - stars: oscillations}}

\maketitle

\section{Introduction}

The problem of the excitation of internal gravity waves (hereafter IGWs)
in stably stratified media is often studied
in connection with the possible detection of solar $g$-modes
(see e.g.\ the latest attempt of $g$-mode detection in the GOLF data by
Gabriel et al.\ 2002).
As an example, two- and three-dimensional numerical simulations
of penetrative convection have shown that it is possible to excite
such waves below the convection zone from the penetration of
strong downward plumes into the stable radiative zone located below
(Hurlburt et al.\ 1986; Hurlburt et al.\ 1994; Brandenburg et al.\ 1996;
Brummell et al.\ 2002; Dintrans et al.\ 2003).

Furthermore, IGWs play a major role in stellar
evolution as they can transport angular momentum and/or chemical elements
over long distances in the stellar interior. This transport mechanism has
been proposed to explain the quasi-solid rotation profile
of the solar core as revealed by helioseismology (Kumar et al.\ 1999) 
as well as the lithium depletion observed in low mass stars (Talon \& Charbonnel
1998). However, the correct amount of IGWs excited by, say, penetrative
convection remains still unknown as numerical studies were essentially
led from a qualitative point of view, so that the mode amplitudes
were not deduced.

The main goal of this paper is to show that it is possible to determine
quantitatively the amplitudes of gravity waves propagating in a stable zone
of a numerical simulation using the anelastic subspace. This subspace is
built from the solutions of the associated anelastic linear eigenvalue
problem for the perturbations. The projection of the simulated velocity
field onto this basis yields the IGW amplitudes. We
will present here the application of this method to the simple problem
of $g$-mode oscillations in an isothermal atmosphere. The dynamics
of such an atmosphere is well known (Brandenburg 1988)
so that this problem is ideal to illustrate and test the validity of our method,
before applying it to the more difficult problem of
IGWs excited by penetrative convection, where the atmosphere is in
general non-isothermal.

After presenting the hydrodynamical model describing our isothermal
atmosphere
(Sect.\ \ref{themodel}), we apply two classical and widely used methods
to detect the $g$-modes excited by an oscillating entropy bubble and
show their limitations (Sect.\ \ref{classic}). Next, we introduce our new
method based on the anelastic subspace and give the analytic forms we
found for both the eigenfrequencies and eigenvectors of the anelastic
set of linear equations for the perturbations (Sect.\ \ref{newmethod}).
We then apply this formalism to the same simulation used to test the two
classical methods discussed in Sect.\ \ref{classic} and show that we now have access to
both the spectrum and amplitudes of the IGWs
(Sect.\ \ref{results}). In particular, we illustrate the
usefulness of our method by studying the influence of
the box geometry on the mode amplitudes. Finally, we conclude in
Sect.\ \ref{conclu} by introducing the next step of this work, which
will consist of the application of the anelastic subspace method to the
detection of IGWs that are stochastically excited by penetrating convective plumes.

\section{The hydrodynamical model}
\label{themodel}

\subsection{The basic equations}

We consider the two-dimensional propagation of internal gravity waves
in an isothermal atmosphere consisting of a layer of depth $d$
of a perfect gas at temperature $T_0$ embedded between two horizontal
impenetrable boundaries. The fluid is non-rotating and stratified with
constant gravity and its properties like its kinematic viscosity, $\nu$,
and specific heats, $c_p$ and $c_v$, are assumed to be constant (with
$\gamma=c_p/c_v=5/3$ for a monatomic gas). Initially, the layer is in
hydrostatic equilibrium such that its density is given by

\beq
\rho_0 (z) = \rho_{\rm top} \exp(z/H),
\eeqn{isoth}
where $z$ is depth, i.e.\ it points downward in the same direction
as gravity $\vec{g}$, $z=0$ corresponds to the top of the box,
$H=c_{\rm s}^2/\gamma g$ is the pressure scale
height, $c_{\rm s}=\sqrt{\gamma R_\ast T_0}=\hbox{const}$ denotes the speed
of sound ($R_\ast$ being the gas constant) and $N^2 =( 1-1/\gamma)
g/H=\hbox{const}$ is the square of the \BV frequency.

We choose the depth of the layer $d$ as the length unit, $\sqrt{d/g}$ as 
the time unit 
and $\rho_{\rm top}$ and
$T_0$ as the units of density and temperature, respectively. The evolution
of this layer is governed by the equations of conservation of mass,
momentum and energy

\begin{equation}
\left\{ \begin{array}{l}
\displaystyle \frac{D \Ln \rho}{Dt} = -\Div \vu, \\ \\
\displaystyle \frac{D \vec{u}}{Dt} = - (\gamma-1) \lp \na e + e \na
\Ln \rho \rp + \vg + \frac{1}{\rho} \na \cdot (2\rho \nu \vec{\sf S}), \\ \\ 
\disp \frac{De}{Dt} = - (\gamma-1) e \Div \vu + 2\nu \vec{\sf S}^2,
\end{array} \right. 
\label{syst1}
\end{equation}
where $D/Dt = \partial / \partial t + \vec{u} \cdot \vec{\nabla}$
is the total derivative, $\rho$ the density, $\vec{u}$ the velocity,
$e$ the internal energy with $P=(\gamma-1)\rho e$, and
$\vec{\sf S}$ is the trace-free rate of strain tensor, given by

\[
{\sf S}_{ij} = {1\over2}\left(\frac{\partial u_i}{\partial x_j} +
\frac{\partial u_j}{\partial x_i} - \frac{2}{3} \delta_{ij} \vec{\nabla}
\cdot \vec{u}\right).
\]

\subsection{Boundary conditions}

When imposing the boundary conditions, we assume the top and bottom
surfaces to be stress-free with the fixed temperature (or,
equivalently, internal energy), that is

\[
\dz{u_x} = \dz{u_z} = u_z = 0 \ltex{and} e=e_0 \ltex{on} z=0,1.
\]
In the horizontal direction, we shall impose periodic conditions for all
fields.

\begin{figure}
\centerline{\includegraphics[width=8cm]{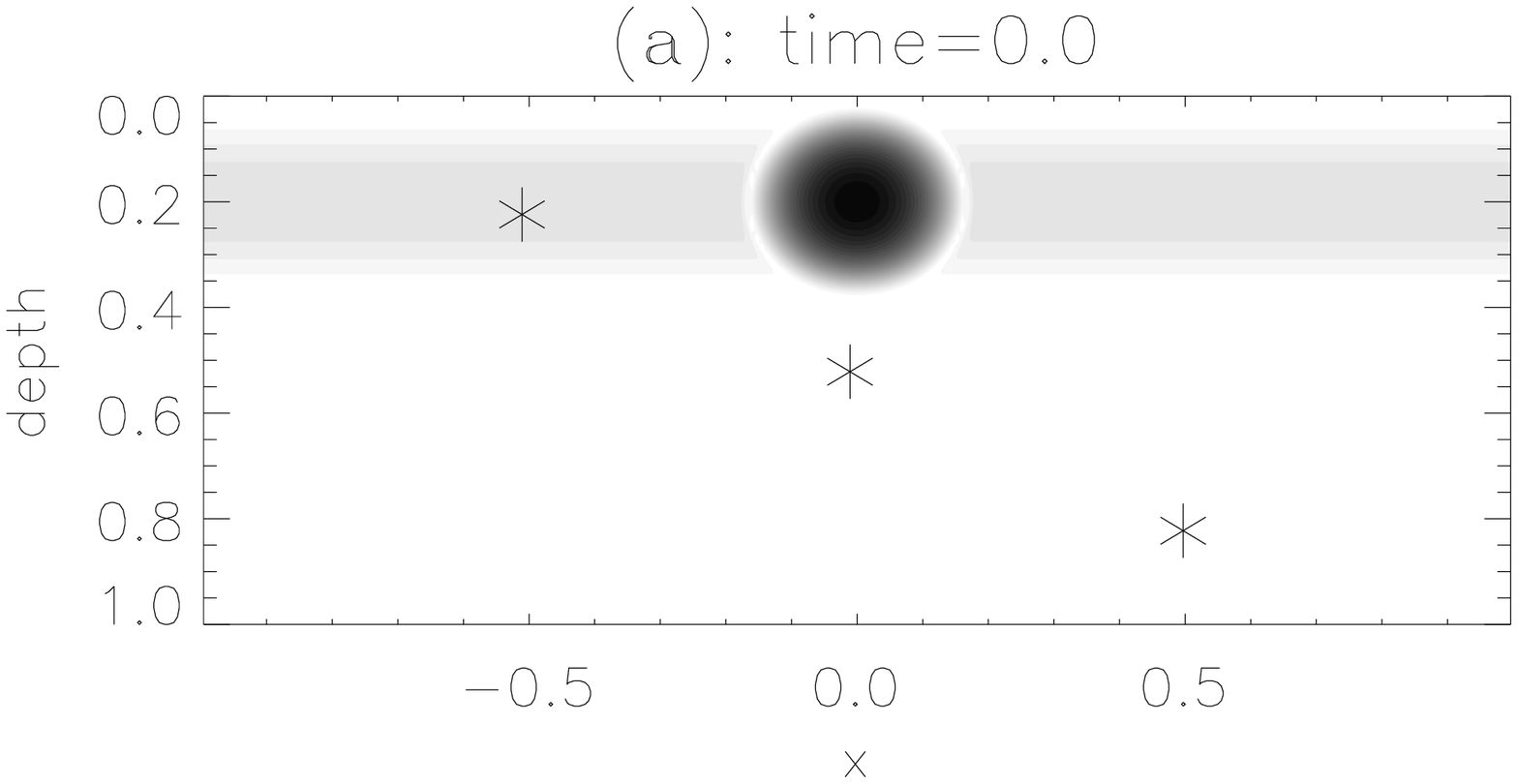}}
\centerline{\includegraphics[width=8cm]{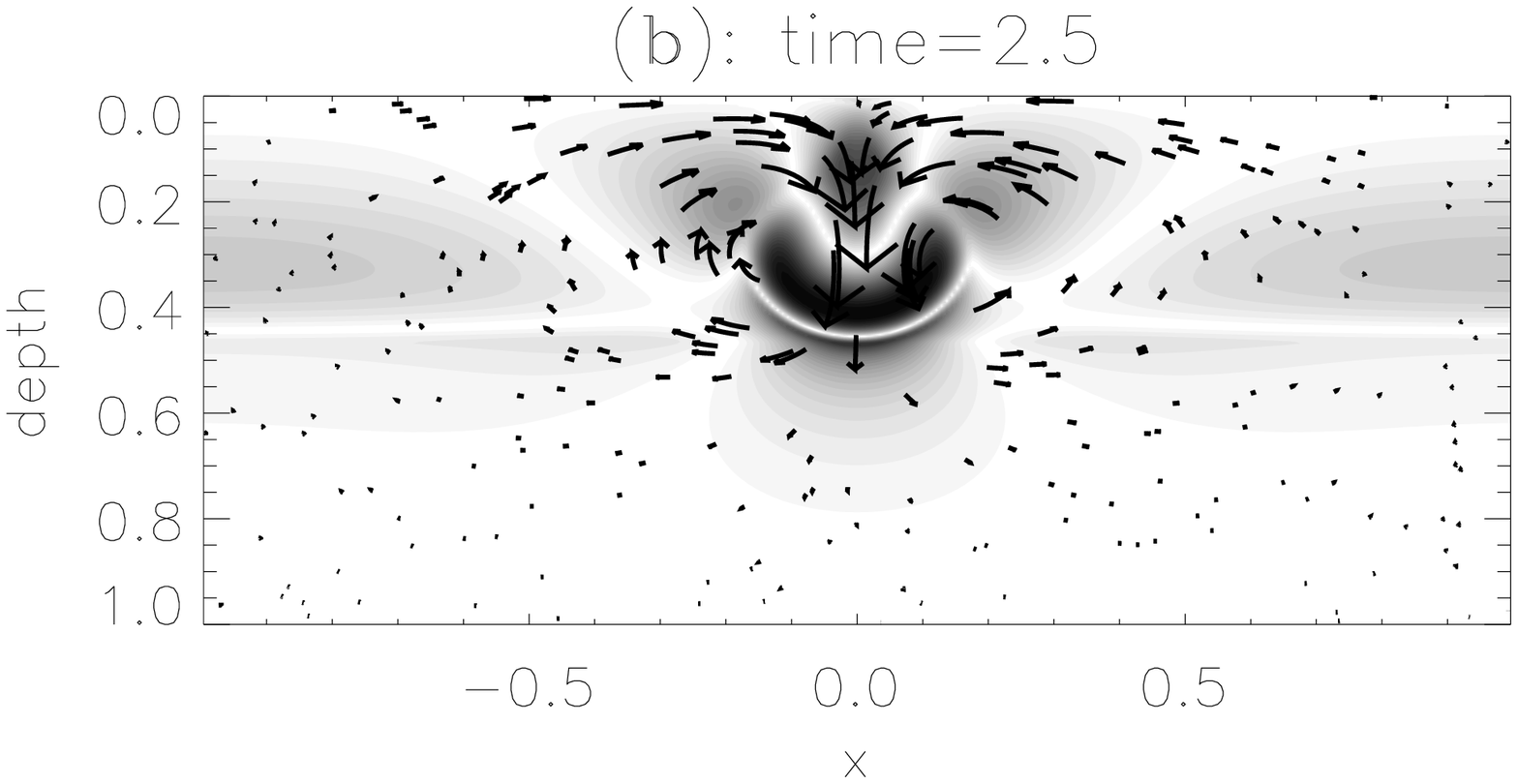}}
\centerline{\includegraphics[width=8cm]{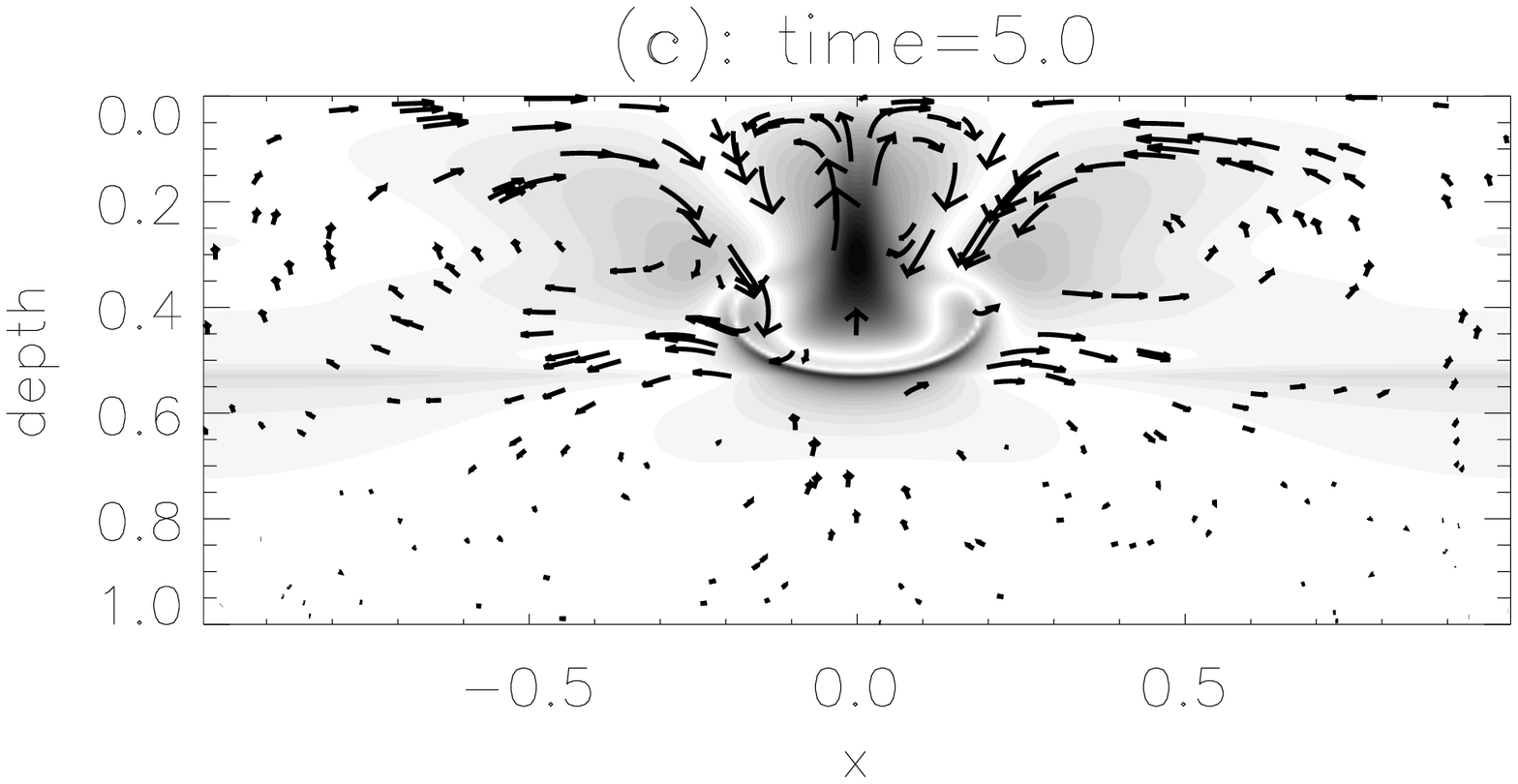}}

 \caption[]{Velocity field superimposed on a grey scale representation
 of the entropy perturbation for a two-dimensional simulation of an entropy
 bubble oscillating in an isothermal atmosphere.
 The asterisks shown in panel (a) are used in
 Fig.~\ref{M1}a.}

\label{sbubb} 
\end{figure}

\subsection{Numerics}

The fully nonlinear set of equations \eq{syst1} is solved using the
hydrodynamical code described in Nordlund \& Stein (1990) and
Brandenburg et al.\ (1996). The time
advance is explicit and uses a third order Hyman scheme (Hyman 1979).
All spatial derivatives are calculated using sixth order
compact finite differences (Lele 1992). The two-dimensional simulations presented
here were done with the same resolution $256\times300$ (i.e.\ 256 zones
in the horizontal direction and 300 in the vertical) and aspect
ratios, $A\equiv L_z/L_x$, between 2 and 6. Here, $L_x$ and $L_z$ denote
the domain size in the horizontal and vertical directions, respectively.
Moreover, the
dimensionless sound speed and the gravitational acceleration
are also the same for all
runs and set equal to $c_{\rm s}=g=1$. As a consequence, the dimensionless
pressure scale height is $H=3/5$ (i.e.\ the box extends over $1.66H$ in the
vertical direction so that the density contrast between bottom and top
is around $\rho_{\rm bot}/\rho_{\rm top}\simeq5.3$), while the \BV frequency
is $N=\sqrt{2/3}\simeq 0.82$. Finally, the dimensionless kinematic
viscosity $\nu$ has been set equal to $\nu=10^{-3}$ in all
simulations, except in the study of the influence of the viscosity
on the mode damping rates (Sect.~\ref{damping_rate}) where we
also considered smaller values down to $\nu=5\times 10^{-4}$.

As initial condition to excite IGWs in the layer, we choose the
perturbation to be an entropy bubble (Brandenburg 1988). For a given
entropy perturbation, it is indeed easy to find the approximate position
of the point around which the bubble will begin to oscillate
and thus generate IGWs. In dimensionless units we have

\[
N^2 = - \dnz{s'} \simeq - \frac{\Delta s'}{\Delta z},
\quad\mbox{so}\quad
\Delta z \simeq - \frac{\Delta s'}{N^2} \simeq - \frac{3}{2} \Delta s',
\]
where $s' \equiv s/c_p = (1/\gamma)\Ln P - \Ln \rho$ is the dimensionless entropy.
All simulations start with an initial (negative) entropy perturbation
$\Delta s'=-0.2$ located at $z=0.2$ so that $\Delta z=0.3$. This
cold bubble then falls down to the middle of the layer, $z=0.5$, and begins
to oscillate around this point, generating IGWs. We note that the entropy perturbation
is done at constant pressure $\Delta P = 0$, i.e.\
$\Delta \ln\rho=-\Delta s'$.

Figure \ref{sbubb} shows an example of such a two-dimensional hydrodynamical simulation 
where
the velocity field has been superimposed onto a grey scale representation
of the entropy. At
$t=0$, the negative entropy perturbation looks like a bubble at $x=0$
and $z=0.2$ (Fig. \ref{sbubb}a) and no velocity field is present. Under
the effect of gravity, the bubble begins to descend (Fig. \ref{sbubb}b)
and stabilizes at the depth $z=0.5$. The bubble thus oscillates
around this equilibrium position and IGWs are generated in the whole domain
(Fig. \ref{sbubb}c). The question is what is the spectrum and amplitude
of the internal waves excited by this oscillating bubble.

\section{Measuring IGWs excited by the oscillating bubble from
classical methods}
\label{classic}

Two techniques are commonly used to measure wave fields propagating
in direct hydrodynamical simulations:

\begin{description}

\item[Method~1 (hereafter M1):] the simplest method consists of firstly,
recording the vertical velocity at a fixed point and, secondly, performing
a Fourier transform of the time series. This method was used for example
by Hurlburt et al.\ (1986) to detect IGWs in a simulation
of penetrative convection.

\item[Method~2 (hereafter M2):] a more precise method consists of, firstly,
taking horizontal Fourier transforms of the vertical mass flux
$\rho w$ for every time step and, secondly, computing power spectra for
the individual time series. Peaks corresponding to acoustic or gravity
oscillations thus appear at a given horizontal wavenumber $k_x$ in the $(z,\omega)$-plane. This
method was used for example by Stein \& Nordlund (1990) to determine
the acoustic modes that are excited in their numerical simulations of
solar granulation.

\end{description}

\begin{figure}
\centerline{\includegraphics[width=8.5cm]{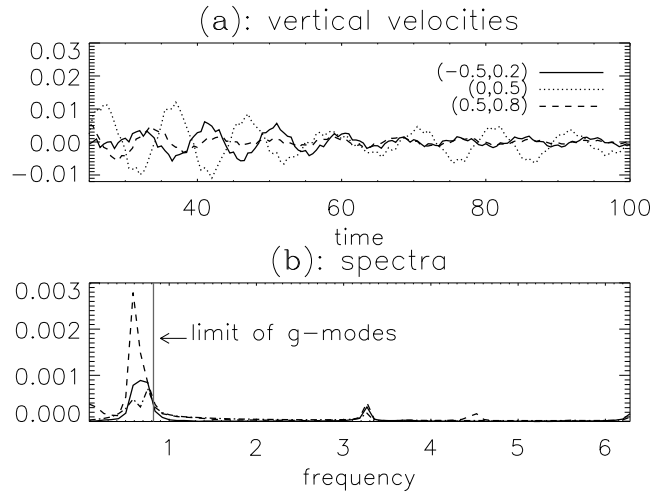}}

 \caption[]{Detection of IGWs using method M1. (a): time evolution of
 the vertical velocity recorded at three different positions in the domain
 (denoted by stars in Fig.~\ref{sbubb}a). (b): corresponding temporal
 power spectra. Peaks below the buoyancy frequency
 $N=0.82$ correspond to internal gravity waves whereas the peak around
 $\omega=\pi$ is the fundamental acoustic mode.}

\label{M1} 
\end{figure}

In Fig.~\ref{M1} we summarize the application of method M1 to detect
the IGWs propagating in the simulation of Fig.~\ref{sbubb}. As
expected for wave propagations, vertical velocities oscillate
approximately periodically around zero at the three different levels
(Fig.~\ref{M1}a). However, when we look at the corresponding temporal
power spectra, only very few modes appear: the fundamental radial
acoustic mode\footnote{It corresponds to sound waves making successive
motions back and forth in the vertical direction with a period $T$
simply given by $T=2/c_{\rm s}=2$, so the pulsation is $\omega=2\pi /T=\pi$.}
and one or two gravity modes below the bounding \BV frequency $N\approx0.82$
(Fig.~\ref{M1}b). We recall that gravity waves cannot propagate along
the vertical direction so that the $g$-modes necessarily correspond to
nonradial modes (e.g.\ Unno et al.\ 1989). Since method M1 does not
take into account the horizontal dependence of the modes, nonradial
$g$-modes with almost the same frequency but different horizontal
wavenumber $k_x$ cannot be distinguished.
This degeneracy in the mode degrees explains why only
very few $g$-modes are captured with this method.
Here and in the following we present the horizontal wavenumbers
in terms of their smallest possible finite value, so that
$k_x=(2\pi/L_x)\ell$ where $\ell$ denotes the mode degree and is an
integer, $\ell=[0,1,2,..]$.

\begin{figure}
\centerline{\includegraphics[width=9cm]{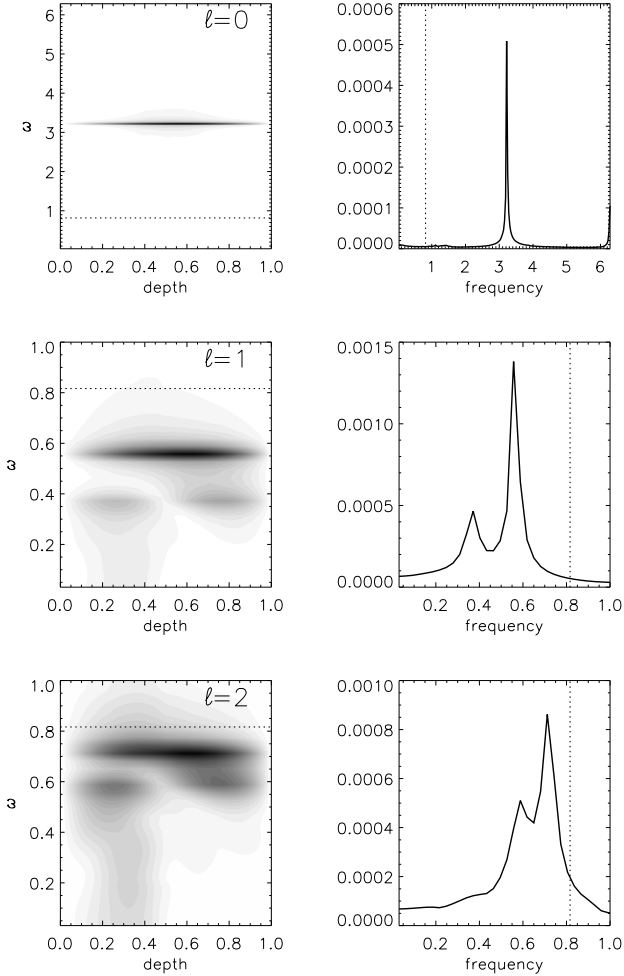}}

 \caption[]{Detection of IGWs using method M2. Left: temporal power spectra
 in the $(z,\omega)$-plane for three different degrees
 $\ell=[0,1,2]$. Right: the resulting spectra after an integration in depth.
 Dotted lines shown in all plots correspond to the bounding frequency
 $\omega=N\approx0.82$, above which $g$-modes cannot exist.}

\label{M2} 
\end{figure}

In Fig.~\ref{M2} we summarize the application of the second method M2
to detect IGWs in the same simulation of Fig.~\ref{M1}. We now have
access to more informations than with method M1. At first, the $\ell=0$
diagram confirms: {\it (i)} that the mode around $\omega=\pi$ is indeed
a radial one and that it corresponds to the fundamental acoustic mode
as no radial node is visible; {\it (ii)} that gravity modes are strictly
nonradial modes as no peaks are visible below the dotted line $\omega=N$.

Second, gravity modes with similar frequency but different degrees $\ell$
are now well separated. As an example, the fundamental $g$-mode at $\ell=1$
and the first overtone at $\ell=2$ have almost the same frequency, $\omega
\simeq 0.6$. Using the $(z,\omega)$-plane for different $\ell$'s
allows us to separate modes and to show that one mode
corresponds to a fundamental one while the other one is a first
overtone [note the two ``bumps'' around $\omega\simeq 0.6$ in the
$(z,\omega)$-diagram for $\ell =2$].

\section{A new method using the anelastic subspace}
\label{newmethod}

The classical methods M1 and M2 presented above have some disadvantages
which render them inappropriate for a careful detection of IGWs in
hydrodynamical simulations:

\begin{itemize}

\item the main disadvantage of method M1 is its $k_x$-degeneracy, that
is, the horizontal dependence of the modes is not taken into account. As
a consequence, we have only access to a rough spectrum where only very few peaks
appear.

\item method M2 takes into account this horizontal dependence so
that more modes are detected. But what about modes amplitudes?
For example, this method does not permit to quantify the amount of kinetic energy
due to IGWs propagating in the box so that only qualitative information, such
as the knowledge of the spectrum of excited $g$-modes, is possible.

\end{itemize}

\noindent Finally, we recall that the goal of this work is to find a
tool allowing us to measure quantitatively IGWs that are stochastically excited by
penetrative convection. In view of this it is clear that methods M1 and
M2 are not well adapted to this problem as Fourier transforms done to
find the spectrum are calculated over the whole simulation. 

Our new method, which eliminates these shortcomings, is inspired by
the work of Bogdan, Cattaneo \& Malagoli (1993, hereafter BCM) who have
developed a tool to measure the {\it acoustic} emission generated in
their simulations of turbulent convection. They extracted the
acoustic field by projecting the convection field onto the acoustic
subspace build from the eigenfunctions of the associated linear
oscillations problem.  Here we are interested in gravity waves rather
than sound waves and this allows us to adopt a simpler procedure than
theirs. Thus, we project our simulation data onto the anelastic
subspace, that is, we filter out the acoustic waves in the linear system
of oscillation equations.  For a time dependence of normal modes of the
form $\exp ({\rm i}\omega t)$, and in the ideal limit $\nu = 0$, the anelastic
set of equations reduces to (Dintrans \& Rieutord 2001; Rieutord \&
Dintrans 2002)

\greq
\disp \omega^2 \vxi = \na \Pi + N^2 \xi_z \ez, \\ \\
\Div (\rho_0 \vxi) = 0, \\ \\
\xi_z = 0 \ltex{for} z=0,1,
\egreqn{anel}
where $\vxi=(\xi_x,\xi_z)$ denotes the Lagrangian displacement
(the associated velocity field being $\vi = {\rm i}\omega \vxi$),
$\Pi=P'/\rho_0$ the eulerian perturbation in reduced pressure and
$\rho_0(z)$ the equilibrium density profile.

Because we adopt periodic boundary conditions in the $x$-direction,
we can seek solutions of the form

\[
\xi_x (x,z) = \xi_x (z) \cos (k_x x) \ltex{and} \xi_z (x,z) =
\xi_z (z)\sin (k_x x),
\]
such that the anelastic system \eq{anel} reads after the elimination of
$\Pi$

\greq
\disp \omega^2 \lp \xi_z - \frac{1}{k_x} \dnz{\xi_x} \rp = N^2 \xi_z, \\ \\
\disp -k_x \xi_x + \dnz{\xi_z} + \dnz{\Ln \rho_0} \xi_z = 0, \\ \\
\xi_z = 0 \ltex{for} z=0,1.
\egreqn{periodic}
This last system can be formally rewritten as a generalized eigenvalue
problem of the form

\beq
{\cal M}_A \vpsi_{\ell n} = \omega^2_{\ell n} {\cal M}_B \vpsi_{\ell n},
\eeqn{eigenv1}
where $\vpsi_{\ell n} = (\xi_x,\xi_z)^T$ is the eigenvector associated with
eigenvalue $\omega^2_{\ell n}$ and ${\cal M}_A$, ${\cal M}_B$ denote two
differential operators.

As shown in Appendix \ref{A1}, the eigenfunctions $\vpsi_{\ell n}$ which are
solutions of the eigenvalue problem \eq{eigenv1} are orthogonal,
that is we have

\beq
\langle \vpsi_{\ell n},\vpsi_{\ell 'n'} \rangle = \int_0^1
\vpsi^\dagger_{\ell n} \cdot \vpsi_{\ell'n'} \rho_0 {\rm d}z = 
\delta_{\ell \ell'}\delta_{nn'},
\eeqn{product}
for normalized eigenfunctions (here the symbol $\dagger$ denotes the
Hermitian conjugate). As a consequence, the set $\vpsi_{\ell n}$
forms an orthogonal basis onto which the simulated velocity field $\vi$
can be projected, i.e.\

\beq
\vi (x,z,t) = \disp \sum_{\ell =1}^{+\infty} \sum_{n=0}^{+ \infty}
\langle \vpsi_{\ell n},\vi \rangle \vpsi_{\ell n} (z)
\left| \begin{array}{l}
\cos (k_x x) \\ \\
\sin (k_x x)
\end{array} \right. + \vi_{\hbox{\scriptsize rest}},
\eeqn{projec}
where the $\ell$-sum begins at $\ell=1$ as no gravity modes propagate
radially, i.e. $k_x =(2\pi/L_x)\ell $ should be non-zero (see
Sect.~\ref{classic}).
In this 
decomposition, the first term
corresponds to the IGW contribution while all other rates, such as the
bubble displacement or the acoustic waves, were collected in a term which
we identify hereafter as ``rest". Finally, following BCM, we identify
the amplitude of each gravity mode by the time-dependent coefficient

\beq
c_{\ell n} (t) = \langle \vpsi_{\ell n},\vi \rangle.
\eeqn{defampli}

In the case of an isothermal atmosphere, for which $\rho_0$ and $N$ are
simply given by \eq{isoth}, analytic expressions for both eigenvalues
and eigenvectors of the anelastic system \eq{periodic} can be found
(see Appendix \ref{A2}) and we give here only the result:

\greq
\disp \omega_{\ell n} = N \lc 1 + \frac{1}{k_x^2} \lp 
\frac{1}{4 H^2} + k^2_z \rp \rc^{-1/2}, \\ \\
\disp \xi^{\ell n}_x = \frac{C}{k_x} \exp(-z/2H) \lc k_z
\cos(k_z z) + \frac{1}{2H} \sin (k_z z) \rc, \\ \\
\xi^{\ell n}_z = C \exp (-z/2H) \sin (k_z z),
\egreqn{theory}
where we have introduced the vertical wavenumber $k_z = (n+1)\pi$. Because of these
analytic expressions, the construction of the anelastic basis used in
the projection \eq{projec} is straightforward. However, in practice, we
do not project directly onto the two-dimensional velocity field but rather its
horizontal Fourier transform at a given $k_x$, so that the procedure is
the following:

\begin{enumerate}

\item given a $k_x$-value, we first deduce the anelastic eigenvalues
$\omega_{\ell n}$ and eigenvectors $\psi_{\ell n}=(\xi^{\ell n}_x,
\xi^{\ell n}_z)^T$
from their analytic expressions \eq{theory}.

\item we build the anelastic subspace at this horizontal wavenumber $k_x$ with, 
say,
the first five anelastic modes $n=[0,4]$.

\item we Fourier transform the simulated velocity field $\vi (x,z,t)$
to obtain the new complex-type field $\hat{\vi}_\ell (z,t)$.

\item we project $\hat{\vi}_\ell (z,t)$ onto the anelastic basis, built
at step 2, and deduce the mode amplitudes \eq{defampli}
for each order $n$.

\end{enumerate}

We will now illustrate the efficiency of our IGW detection method by
measuring IGWs excited in the same simulation used to present methods M1
and M2 in Sect.~\ref{classic}.

\section{Results}
\label{results}

\begin{figure}
\centerline{\includegraphics[width=8cm]{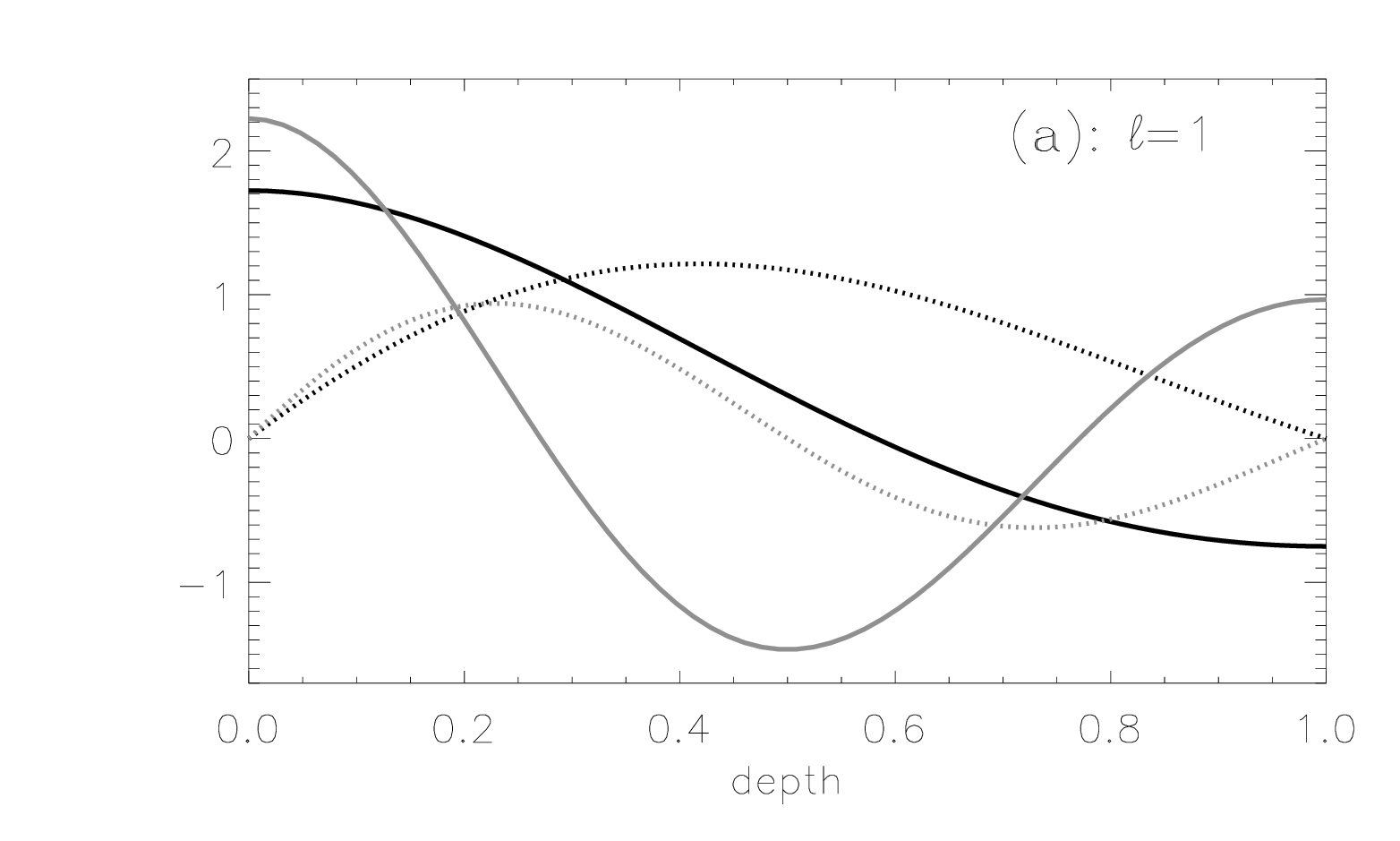}}
\centerline{\includegraphics[width=8cm]{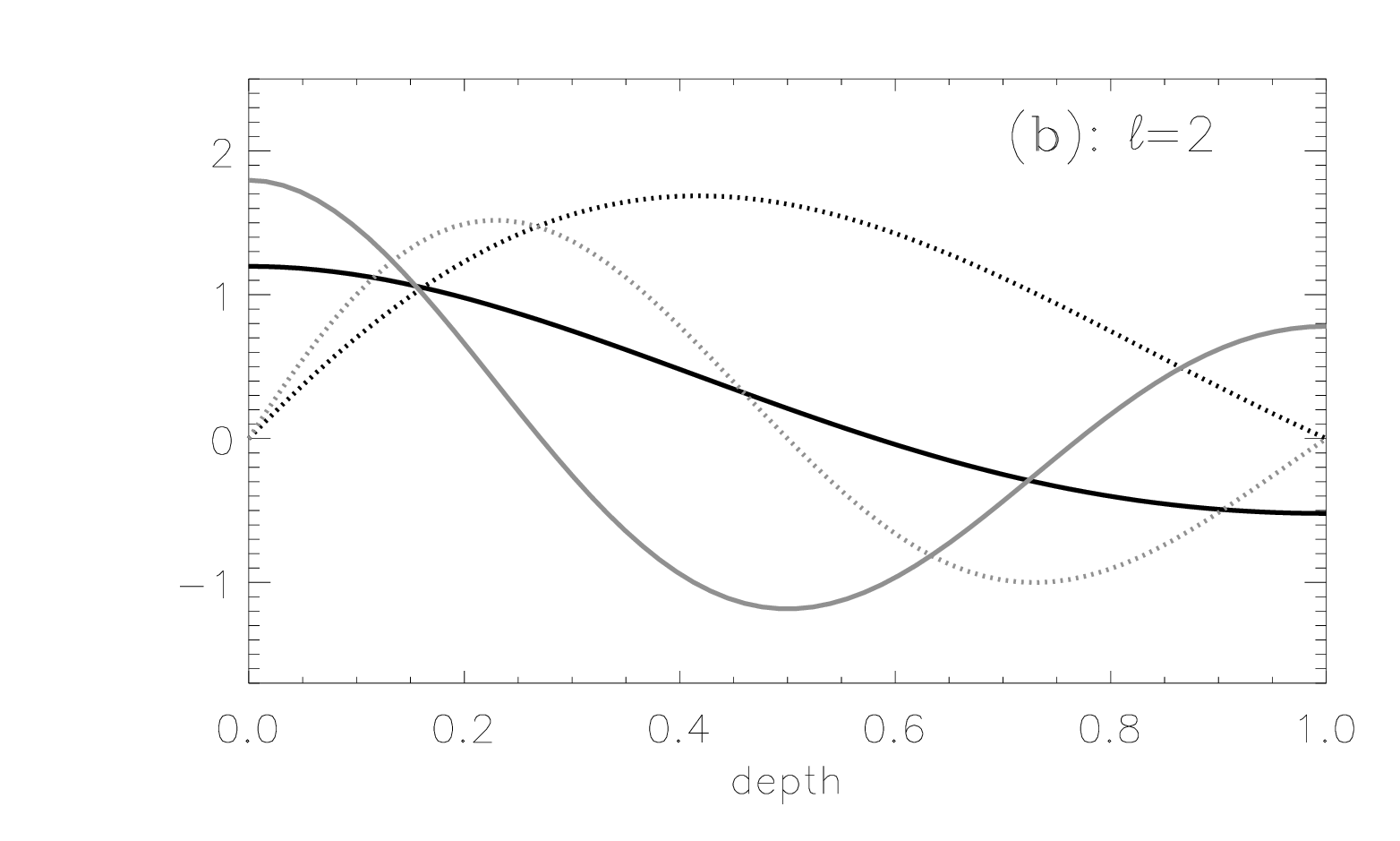}}

 \caption[]{Vertical structure of the fundamental anelastic mode (dark
 line) and first overtone (grey line) at $\ell = 1$ (a) and $\ell = 2$
 (b). Solid lines mark the horizontal displacement $\xi_x$ while
 dot-dashed lines are for the vertical one $\xi_z$. Eigenfunctions have
 been normalized by imposing $\int_0^1 |\psi_{\ell n}|^2\rho_0 dz=1$.}

\label{vecpanel} 
\end{figure}

\subsection{The anelastic eigenvectors: comparisons with the simulation
profiles}

Figure \ref{vecpanel} shows four anelastic modes at $\ell =1$ (a) and
$\ell =2$ (b) with $n=0$ (dark lines) and $n=1$ (grey lines). We remark that
anelastic eigenfunctions hardly change with the $\ell$-values as $\xi_z$
does not explicitly depend on $k_x$, while only the $\xi_x$-amplitude
involves $k_x$. As a consequence, it is the ratio $\xi_z / \xi_x
\propto k_x$ which makes sense so that $\xi_z$ becomes more and more
important compared to $\xi_x$ with increasing $k_x$, as observed in
Fig.~\ref{vecpanel} when comparing solid to dot-dashed lines for the
same order $n$. In the case of method M2,
the use of the vertical mass flux $\rho_0 v_z$ (instead of, say,
$\rho_0 v_x$) is justified {\it a posteriori}
as a good proxy for detecting IGWs propagating in the
simulation (see Fig.~\ref{M2}).

\begin{figure}
\centerline{\includegraphics[width=8cm]{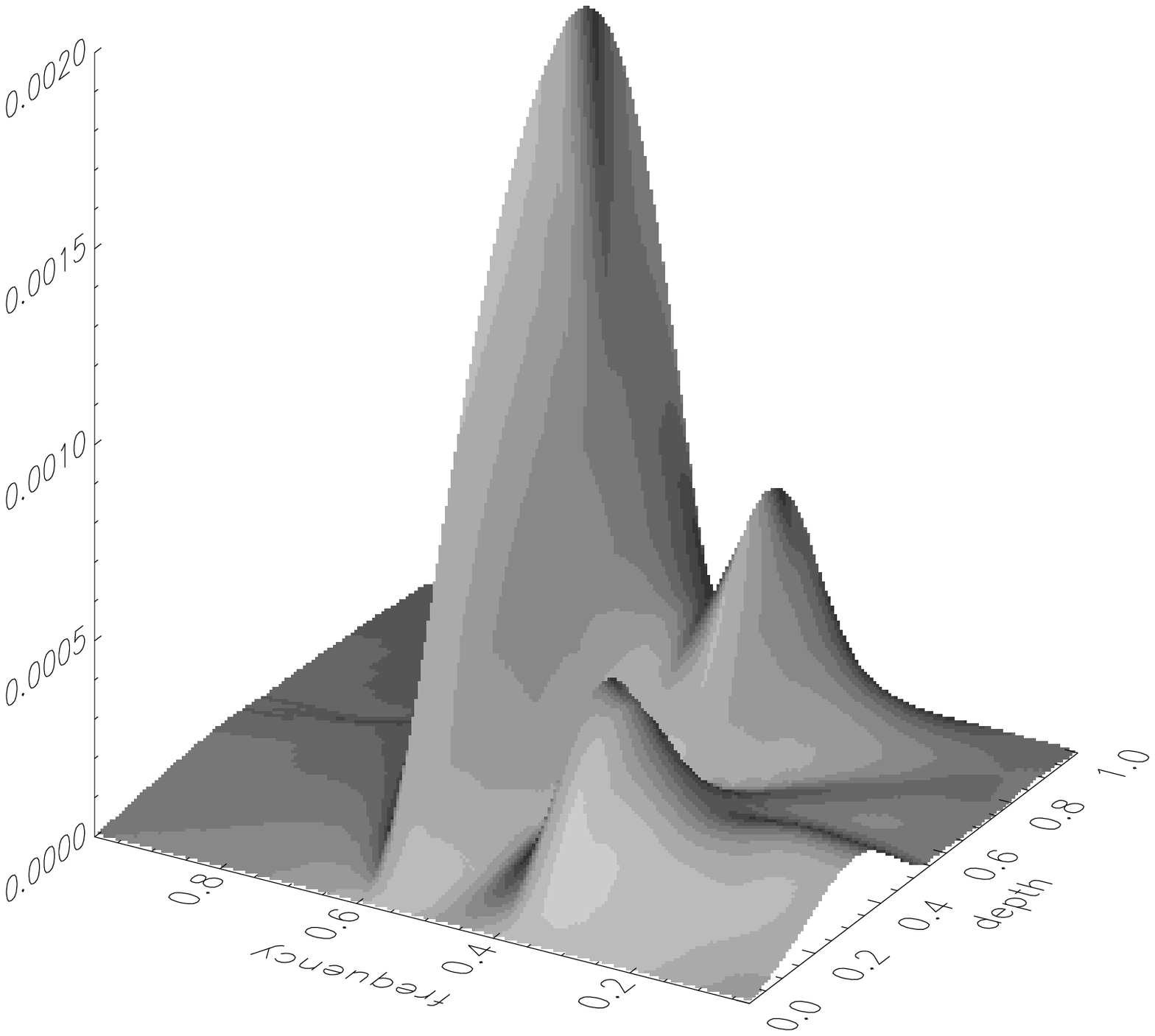}}
\centerline{\includegraphics[width=8cm]{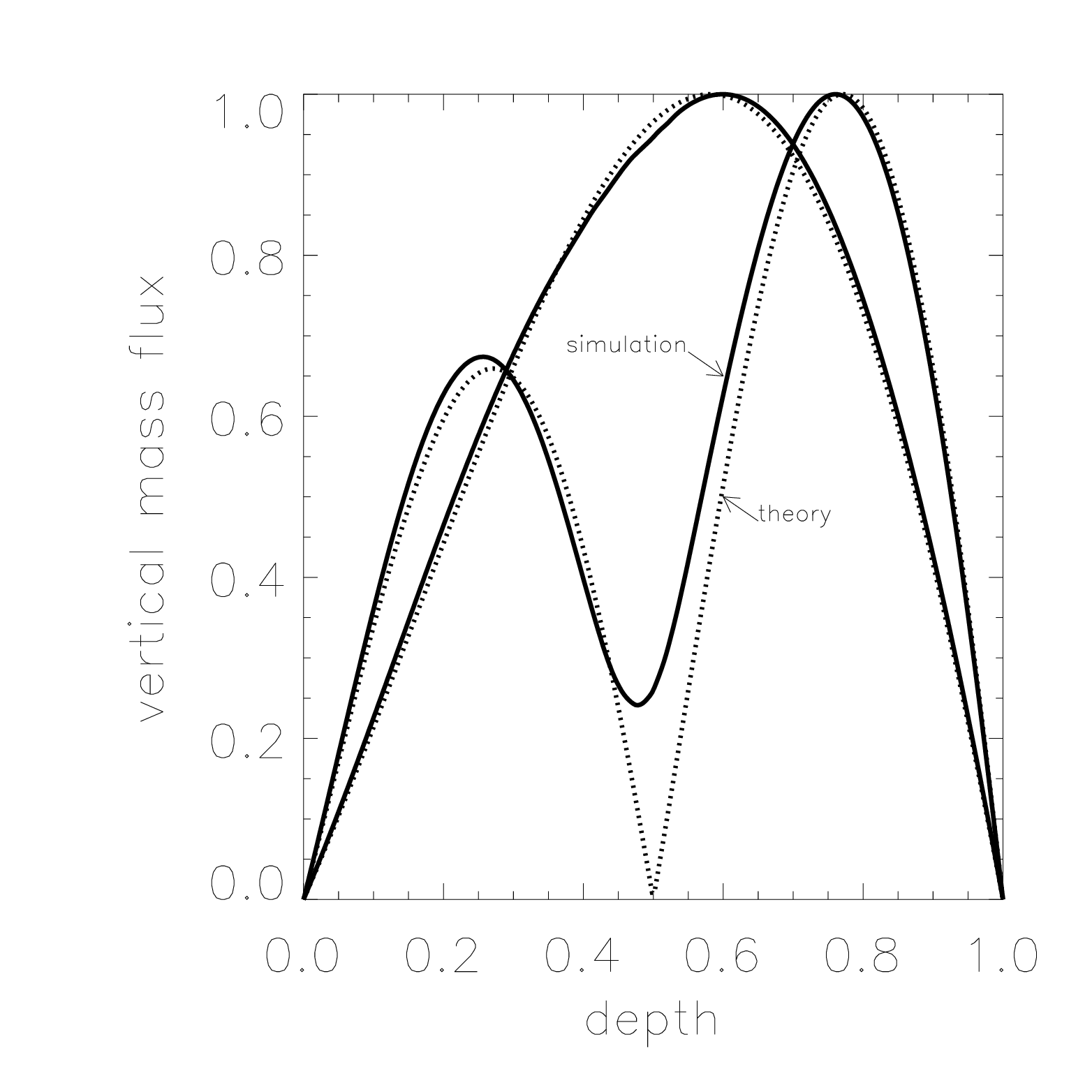}}

 \caption{Upper: same as Fig.~\ref{M2} at $\ell=1$, except that amplitudes
 have been restored using third direction. Lower: comparisons between
 the normalized vertical mass flux $\rho_0 v_z$ profiles deduced from
 the upper power spectra (solid lines) with the ones obtained from the
 anelastic modes shown in Fig.~\ref{vecpanel}a (dotted lines).}

\label{surf}
\end{figure}

Once the anelastic eigenfunctions are known, one can compare the corresponding
vertical mass flux $\rho_0 v_z$ with the ones deduced from the power
spectra used in method M2. Indeed, the peaks appearing at a given
frequency in Fig.~\ref{M2} correspond to $g$-modes so that their vertical
profiles can be related to the eigenfunctions of the linear modes. This
was done by BCM for the case of acoustic modes propagating
in their convection simulations.

The upper panel of Fig.~\ref{surf} shows the depth dependence of the
two peaks appearing at $\ell=1$ in Fig.~\ref{M2}, which correspond to
the $n=0$ ($\omega_{10} = 0.568$) and $n=1$ ($\omega_{11} = 0.363$)
modes, respectively. We thus recover what BCM called the ``shark
fin'' profiles, that is, the power at any depth is located around
the theoretical anelastic frequency. Taking a mean profile around
eigenfrequencies $\omega_{10}$ and $\omega_{11}$ allows us to compare
the vertical mass flux observed in the simulation with those calculated
from the linear anelastic modes. The lower panel of Fig.~\ref{surf}
shows that the agreement between the two profiles is quite remarkable,
meaning that the anelastic modes reproduce well the long-period dynamics of the
oscillating isothermal atmosphere.

\subsection{Time evolution of the mode amplitudes $c_{\ell n} (t)$}
\label{damping_rate}

\begin{figure}
\centerline{\includegraphics[width=8cm]{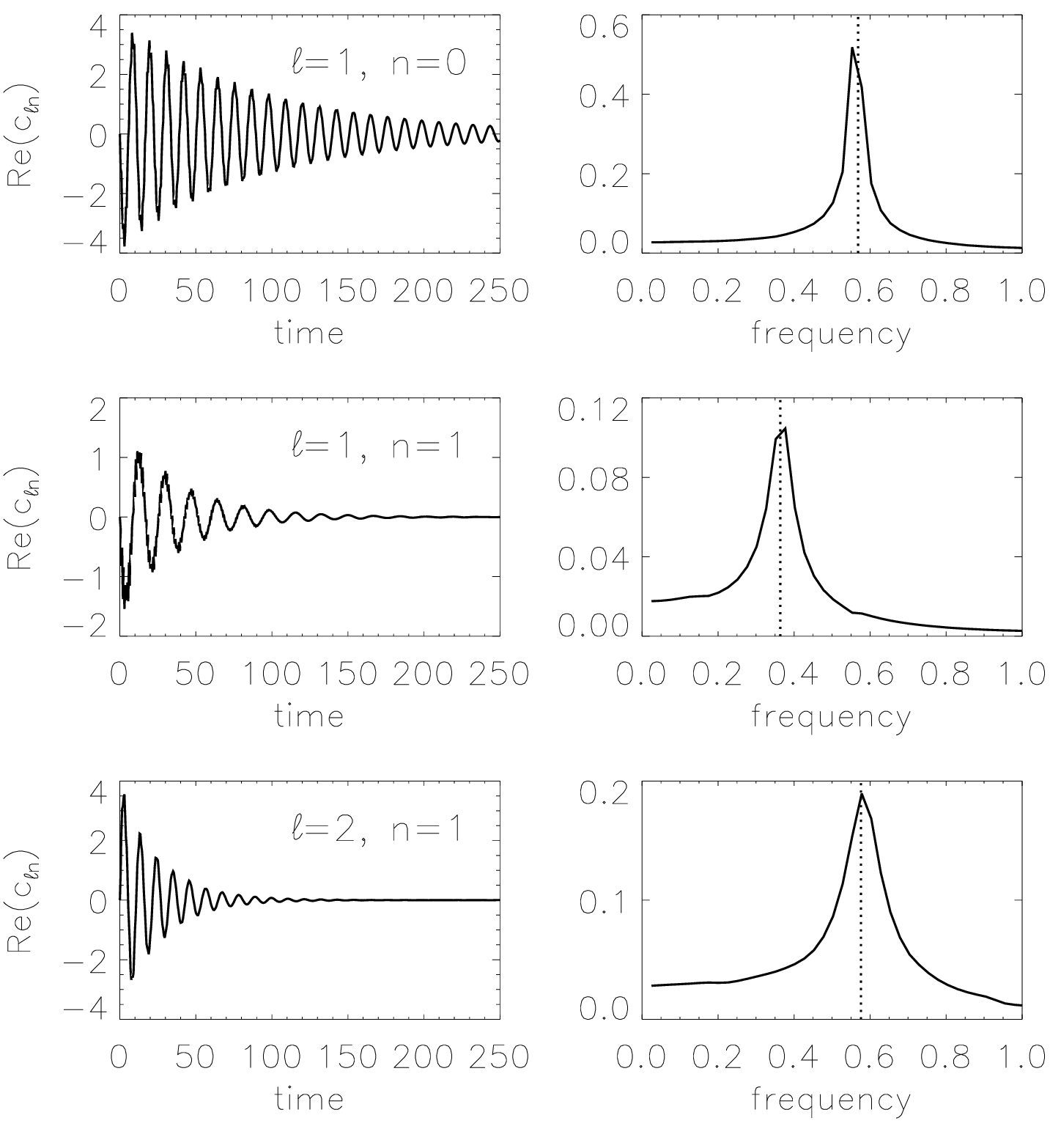}}
 \caption[]{Left: real parts of the complex mode amplitude $c_{\ell n}
 (t)$ for the three modes shown in Fig.~\ref{vecpanel}. Right:
 corresponding temporal power spectra. Dotted lines mark the location
 of the anelastic eigenmodes $\omega_{10}=0.568,\ \omega_{11}=0.363$
 and $\omega_{21}=0.575$, respectively. Amplitudes on each plot have
 been multiplied by one thousand.}
\label{reA} 
\end{figure}

The left hand panel of Fig.~\ref{reA} shows the real parts of the complex
amplitudes $c_{10},\ c_{11}$ and $c_{21}$, that is,
the result of the projection of the simulation shown in Fig.~\ref{sbubb}
onto the three anelastic modes plotted in Fig.~\ref{vecpanel}. By taking
a Fourier transform of these sequences (Fig.~\ref{reA}, right), one obtains
three peaks which agree remarkably well with the theoretical anelastic
frequency given by \eq{theory}, i.e.\ the complex amplitudes $c_{\ell n}$
behave as 

\beq 
c_{\ell n} \propto \exp ({\rm i}\omega_{\ell n} t).
\eeqn{period} 

\noi This relation means that when an eigenmode is excited in
the simulation,
its corresponding displacement is periodic with a period $T_{\ell n} =
2\pi/\omega_{\ell n}$.

\begin{figure}
\centerline{\includegraphics[width=9cm]{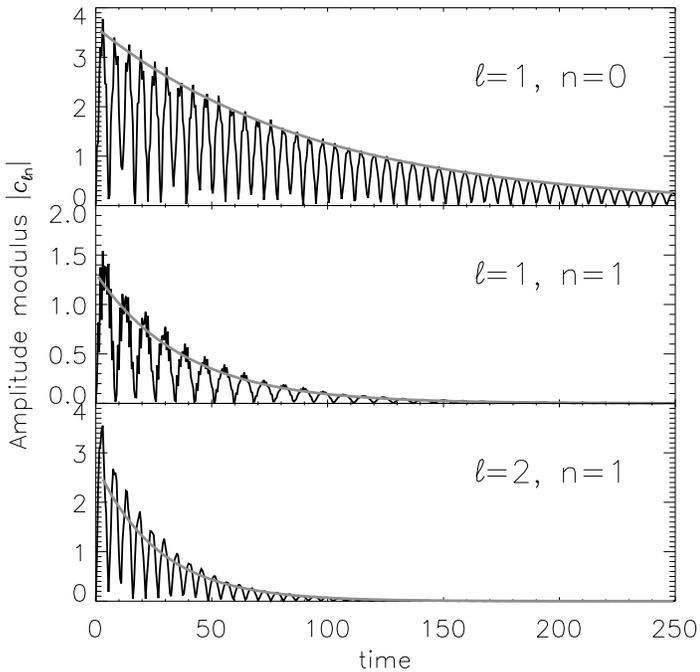}}

 \caption{Time-evolution of the amplitude modulus $|c_{\ell n}|$ for the
 three modes shown in Fig.~\ref{vecpanel}. Grey lines superimposed
 correspond to an exponential decay of the form $|c_{\ell n}| \propto
 \exp(-\alpha t)$.}

\label{decay}
\end{figure}

As we have access to the real and imaginary parts of $c_{\ell n}$, we now
compute the amplitude modulus $|c_{\ell n}|$; Fig.~\ref{decay} shows its time
evolution for the three modes in Fig.~\ref{reA}. A regression analysis
of the curves $\max(|c_{\ell n}|)=f(t)$ allows us to determine the shape of
the envelope and we found that $|c_{\ell n}|$ follows a simple exponential
decay law of the form $\exp (-\alpha t)$. Because of the periodic behaviour
\eq{period}, it means that the mode amplitude behaves like

\beq
c_{\ell n} \propto \exp (-\alpha t) \exp ({\rm i}\,\omega_{\ell n} t),
\eeqn{harm}
where $\alpha $ is a constant which is related to the damping
process, here the viscosity. Using runs with different
kinematic viscosities we have verified that $\alpha  \propto \nu$,
i.e.\ every mode obeys the same law as that
of a linear damped harmonic oscillator. Such a relation between the
damping of the mode and the viscosity is hardly surprising as it can
easily be deduced from work integrals; see Appendix \ref{A3}.

\subsection{Contributions of $g$-modes to the time-averaged kinetic energy}

\begin{figure}
\centerline{\includegraphics[width=9cm]{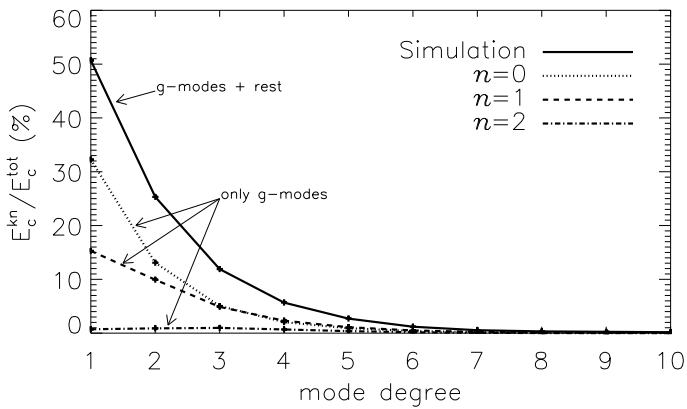}}
  \caption{Dependence of the time-averaged kinetic energy on
  degree $\ell$ and radial order $n$.}
\label{power}
\end{figure}

We will now show that one of the main advantages of the anelastic subspace
resides in the fact that it allows to calculate the contribution of 
every $g$-mode to
the time-averaged kinetic energy.
We demonstrate in Appendix \ref{A4} that a Parseval-Bessel relation
also exists in the anelastic subspace as

\beq
\int_V \rho_0 \vi^2(x,z) \,{\rm d}x\,{\rm d}z = \sum^{+\infty}_{\ell,n}
|c_{\ell n}|^2 + {\rm rest},
\eeqn{pars2}
where, as before, the rest term contains all the contributions which
are not due to IGWs.

This relation illustrates the usefulness of the anelastic subspace when
one wants to perform a detailed study of the IGW excitation. By using
the classical Parseval-Bessel relation applied in Fourier space,
it is already possible to find the amount of kinetic energy contained at
a given wavenumber $k_x$, since

\beq
\int_V \rho_0 \vi^2(x,z) \,{\rm d}x \,{\rm d}z = \int_z
\sum^{+\infty}_{\ell} 
|\hat{\vi}_\ell|^2 \rho_0 \,{\rm d}z.
\eeqn{pars3}
For example, the dependence of the time-averaged kinetic energy on the
degree $\ell$, shown in Fig.~\ref{power}, demonstrates that half 
of the
kinetic energy in the simulation in Fig.~\ref{sbubb} is contained in
the $\ell=1$ degree (the solid line denoted by ``$g$-modes + rest").
Unfortunately, this curve gives no information about the vertical
dependence of this distribution so that contributions coming from
nonradial acoustic modes as well as gravity modes can be present.

The main advantage of our anelastic subspace method is that it lifts this
degeneracy by isolating contributions that come from $g$-modes {\it only}.
Comparing
to its classical form \eq{pars3}, the anelastic Parseval-Bessel relation
\eq{pars2} introduces the radial order $n$ so that one can extract
the contribution of the $g$-mode $(\ell,n)$ in the kinetic energy and then
deduce which modes contribute most to it, meaning the modes that are
the most excited by the oscillating bubble.
This is what we did in Fig.~\ref{power} where we also plotted, for each
$\ell$, the 
contributions coming from
the $n=0$, $n=1$ and $n=2$ $g$-modes. It allows us to show that the
$\ell=1$
contribution is in fact almost entirely composed of the $(1,0)$ and
$(1,1)$ $g$-modes, as they respectively contain around 33\% and 15\%
of the time-averaged kinetic energy of this simulation. This kind of information
is particularly relevant when one deals with the problem of wave
excitation.

\subsection{Effects of the box geometry on modes amplitudes}

In order to assess the restriction imposed by the finite size of the box,
we now study the influence of the box geometry onto the wave excitation.
We therefore perform three runs with three different aspect ratios (2, 4 and 6)
and calculate the $g$-mode contributions to the time-averaged kinetic energy
in each case.

\begin{figure}[h]
\centerline{\includegraphics[width=9cm]{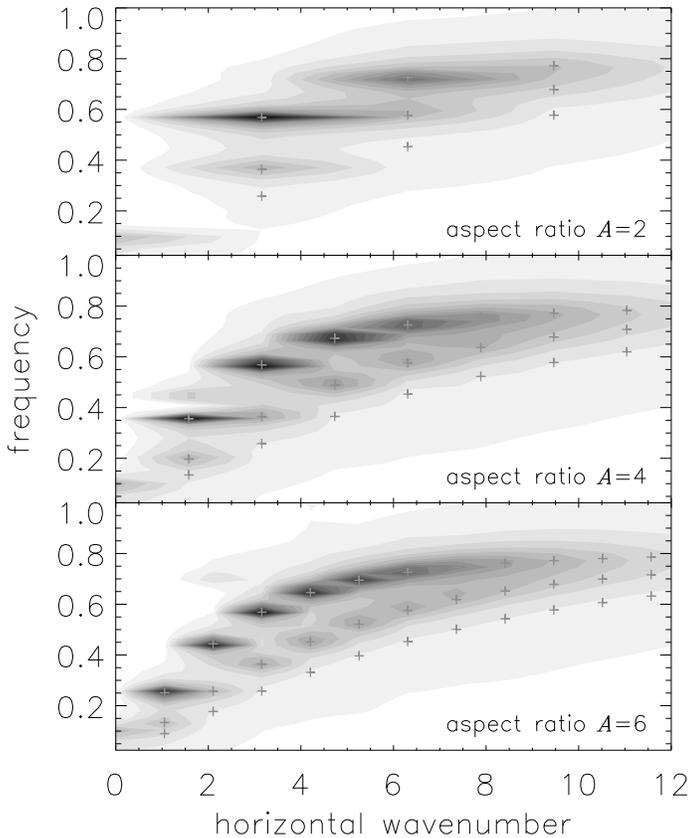}}
 \caption{$(k_x-\omega)$ diagrams which three different aspect ratios (2, 4 and 6).
 Crosses correspond to the anelastic frequencies given by \eq{theory}.}
\label{diagko}
\end{figure}

Figure~\ref{diagko} shows the spectral power in the $(k_x,\omega)$
plane for the three different aspect ratios.
One recovers clearly the well-known dispersion relation for $g$-modes
(e.g.\ Stein \& Leibacher 1974); see also its anelastic counterpart
in Eq.~(\ref{theory}).
One also verifies that
increasing the horizontal extent of the box allows the power to
be distributed among a larger number of horizontal modes. This phenomenon
has also been observed in the problem of $p$-modes excitation by turbulent
convection (Stein \& Nordlund 2001; Nordlund \& Stein 2001). 
Unfortunately, such diagrams are generally not easy to interpret when the
system is not isothermal and the modes stochastically driven; see for
example the three-dimensional simulations of Brandenburg et al.\ (1996)
where the resulting $(k_x,\omega)$ turned out to be very noisy.

\begin{figure}[h]
\centerline{\includegraphics[width=8cm]{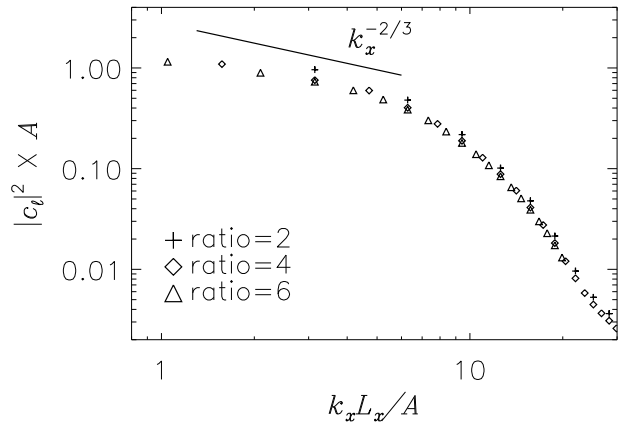}}
  \caption{Distribution of the energy contained in $g$-modes for
  different aspect ratios (2,4 and 6) in a log-log representation.}
\label{distribk}
\end{figure}

Contrasting this with the anelastic subspace method, it is straightforward
to extract at a given wavenumber $k_x$ the energy from modes with different
radial order $n$.
As expected, the spectral distribution of mode energies integrated over
all $n$ collapses onto an universal curve (Fig.~\ref{distribk}).
It turns out that the mode energy scales as $\sim k_x^{-2/3}$ for small
enough values of $k_x$, i.e. for $k_x L_x/A \infapp 10$. However, both
the exponent and the cut-off are not unique and depend on the size of
the initial entropy bubble, as verified by running several simulations
with bubble radii ranging from $R=0.1$ to $R=0.4$. We found that the
larger the bubble, the smaller is the cutoff value of $k_x$, meaning that big
bubbles preferentially generate $g$-modes at large scales. It would be
interesting to investigate in more detail the origin of such behaviour,
but this is clearly beyond the scope of this paper.

\section{Conclusions}
\label{conclu}

We have presented a new and accurate method to measure the internal
gravity waves propagating in a numerical simulation. This is of
prime importance when one deals with the problem of the generation of
IGWs in a stably stratified medium such as the radiative zone of a
star. The knowledge of the vigour of the velocity field in such a zone
is the main input of any theory of wave transport of angular
momentum and/or chemicals in stellar radiative zones
(e.g.\ Schatzman 1993).

However, we have shown that classical methods based only on Fourier
transforms in space and/or time of the simulated velocity field do not
allow a quantitative determination of the mode amplitudes. The reason is
mainly that the contribution of the propagating $g$-modes to the total
energy in the simulation is diluted among the other ones due to $p$-modes
and turbulent motions.

Our method is based on the projection of the simulated velocity field onto
the anelastic subspace built from the solutions of the linear problem
for the perturbations. Once the resulting time-dependent coefficients of
the velocity field are computed, one can deduce the amplitudes
of every $g$-mode and measure their kinetic energy
at every timestep of the simulation.

We have applied this formalism to the measurement of IGWs excited by an
entropy bubble oscillating in an isothermal atmosphere. We have shown that
the mode amplitudes follow the same law than those of a linear
damped harmonic oscillator, that is, the periodic amplitudes decrease
exponentially with time, with a time scale that is simply proportional to
the inverse of the viscosity; see Eq.~(\ref{harm}). By using a 
Parseval-Bessel
relation valid in the anelastic subspace, we have deduced the contribution
of every $g$-mode to the time-averaged kinetic energy and have shown
that large-scale motions are mainly composed of $n=0$ and $n=1$ 
modes.
Such a result is in general impossible to obtain in a more complicated
settings using a classical
$(k_x-\omega)$ diagram.

The fact that the mode amplitudes in the anelastic basis follow the same
law than those of a linear damped oscillator is important in
the case of a stochastic excitation of IGWs, such as the one encountered
when one deals with numerical simulations of penetrative convection.
Indeed, in this case the mode
amplitude $c_{\ell n}$ evolves either chaotically around zero when the mode is
not excited, or in a periodic fashion following Eq.~\eq{harm} when the mode is excited.
By computing time-frequency diagrams of that time-dependent amplitude, one
can extract the time intervals during which the mode has been
really excited. In a forthcoming paper, we will present the application
of the anelastic subspace method we tested here to the problem of IGWs
stochastically excited by penetrative convection (Dintrans et al.\ 2004).

\acknowledgements

This work has been supported by the European Commission under Marie-Curie
grant no.~HPMF-CT-1999-00411.
Calculations were carried out on the CalMip machine of the
`Centre Interuniversitaire de Toulouse' (CICT) which is gratefully
acknowledged. We also thank Michel Rieutord and the referee for useful
comments.

\appendix

\section{Orthogonality of the anelastic eigenfunctions}
\label{A1}

We start from the complete set of equations written for two eigenmodes
$(\omega_1,\vxi_1)$ and $(\omega_2,\vxi_2)$ (Dintrans \& Rieutord 2001)

\greq
\omega_1^2 \rho_0 \vxi_1 = \rho_0 \na \Pi_1 + N^2 \rho_0 \xi^{(1)}_z \ez,
\\ \\
\omega_2^2 \rho_0 \vxi_2 = \rho_0 \na \Pi_2 + N^2 \rho_0 \xi^{(2)}_z \ez,
\\ \\
\Div (\rho_0 \vxi_1) = \Div (\rho_0 \vxi_2) = 0, \\ \\
\xi^{(1)}_z = \xi^{(2)}_z = 0 \ltex{for} z=0,1,

\egreq
where $\Pi=P'/\rho_0$ denotes the eulerian fluctuation of the reduced
pressure. Multiplying the $\vxi_1$-equation by $\vxi_2$ and the
$\vxi_2$-equation by $\vxi_1$, and subtracting the resulting two equations
leads to

\beq
(\omega_1^2 - \omega_2^2) \rho_0 \vxi_1\cdot\vxi_2 = \Div (\Pi_1 \rho_0
\vxi_2 ) - \Div (\Pi_2 \rho_0 \vxi_1 ),
\eeq
where we have used the relation $\rho_0 \vxi_2 \cdot \na \Pi_1 = \Div (\Pi_1
\rho_0 \vxi_2) - \Pi_1 \Div (\rho_0 \vxi_2) = \Div (\Pi_1 \rho_0 \vxi_2)$,
as the anelastic approximation implies $\Div (\rho_0 \vxi_2)=0$.

Finally, we integrate this last equation over the volume and use the
Green-Ostrogradsky theorem to obtain

\beq
(\omega_1^2 - \omega_2^2) \int_V \rho_0 \vxi_1\cdot\vxi_2 {\rm d}V = 
\int_S \Pi_1 \rho_0 \vxi_2 \cdot {\rm d} \vec{S} - \int_S \Pi_2 \rho_0 \vxi_1
\cdot {\rm d} \vec{S}.
\eeq

The applying of periodic horizontal boundary conditions and the condition
$\xi_z = 0$ at $z=0$ and $z=1$ leads to the vanishing of the two right
hand side terms. The uniqueness of the solutions ensures that $\omega_1
\neq \omega_2$ so that we conclude that the eigenvectors are orthogonal
to each other, i.e.\

\beq
\int_V \rho_0 \vxi_1\cdot\vxi_2 {\rm d}V = 0.
\eeq

\section{Anelastic oscillations of an isothermal atmosphere}
\label{A2}

\subsection{Derivation of the anelastic eigenfrequencies and
eigenfunctions}

We start from the anelastic set of equations \eq{periodic}

\greq
\disp \frac{N^2}{\omega^2} \xi_z = \xi_z - \frac{1}{k_x} 
\dnz{\xi_x}, \\ \\
\disp -k_x \xi_x + \dnz{\xi_z} + \dnz{\Ln \rho_0} \xi_z =0, \\ \\
\xi_z = 0 \ltex{for} z=0,1,
\egreqn{sysanel}
where, for an isothermal atmosphere, $\rho_0(z)$ and $N^2$ are given by
Eq.~\eq{isoth}. By eliminating $\xi_x$, we arrive at the following
second-order differential equation for $\xi_z$

\greq
\disp \ddnz{\xi_z} + \frac{1}{H} \dnz{\xi_z} - k_x^2 \lp 1 - 
\frac{N^2}{\omega^2} \rp \xi_z =0, \\ \\
\xi_z(0) = \xi_z(1) = 0.
\egreqn{eq_anel}
All coefficients of this differential equation being constants, we can
seek solutions of exponential type $\xi_z \propto \exp(rz)$ to obtain
the characteristic equation

\beq
r^2 + \frac{r}{H} - k_x^2 \lp 1 - \frac{N^2}{\omega^2} \rp = 0,
\eeq
whose general solution is of the form

\beq
r = \frac{1}{2} \lp - \frac{1}{H} \pm{\rm i}\sqrt{\Delta} \rp \ltex{with}
\Delta = - \frac{1}{H^2} - 4k_x^2 \lp 1 - \frac{N^2}{\omega^2} \rp,
\eeqn{delta}
leading to the following solution for $\xi_z$

\beq
\xi_z (z) = \exp(- z/2H) \lc A \cos\lp \frac{\sqrt{\Delta}}{2} z\rp +
B \sin \lp \frac{\sqrt{\Delta}}{2} z\rp \rc.
\eeq
To find the quantization rule for $\Delta$, we
consider the boundary conditions on $\xi_z$,

\greq
\xi_z (0) = 0 \Rightarrow A=0, \\ \\
\disp \xi_z(1) = 0 \Rightarrow \frac{\sqrt{\Delta_n}}{2} = (n+1) \pi 
\ltex{or} \Delta_n = 4(n+1)^2 \pi^2,
\egreq
where $n=[0,1,2\dots]$ is the radial order of the mode. [The presence of the
$n+1$ factor in the quantization rule is consistent with the fact that $\Delta$
is strictly non-zero.] We recall that the definition \eq{delta}
for $\Delta$ allows us to deduce an expression for the anelastic
eigenfrequencies,

\beq
\omega_{\ell n} = N \lc 1 + \frac{1}{k_x^2} \lp 
\frac{1}{4 H^2} + k_z^2 \rp \rc^{-1/2} \mbox{with
$k_z = (n+1)\pi$},
\eeqn{wanel}
and their associated eigenfunctions

\greq
\disp \xi^{\ell n}_x = \frac{C}{k_x} \exp(-z/2H) \lc k_z
\cos(k_z z) + \frac{1}{2H} \sin (k_z z) \rc, \\ \\
\xi^{\ell n}_z = C \exp (-z/2H) \sin (k_z z).
\egreq
The remarkable fact to be noted here is that the vertical lagrangian
displacement $\xi^{\ell n}_z$ does not depend on $k_x$ (the horizontal
wavenumber) but only on the mode order $n$ (see Fig.~\ref{vecpanel}).

\subsection{Comparisons with the complete case}
\label{shooting}

\begin{table}
\caption{Exact eigenfrequencies $\omega$ for the first five $g$-modes of
degree $\ell = 1$, corresponding to an horizontal wavenumber
$k_x = 2\pi/L_x =\pi$ (here $L_x=2$), in the complete 
case, in the anelastic approximation, and 
the corresponding relative errors.}
\begin{tabular}{cccc}
\hline
Order $n$ & $\omega$ & $\omega_{\hbox{\scriptsize anel}} 
$ & rel. error \\ \hline
0  &  0.572 054  &  0.567 455  &  0.804\%  \\
1  &  0.363 084  &  0.362 606  &  0.132\%  \\
2  &  0.257 381  &  0.257 295  &  0.033\%  \\
3  &  0.197 644  &  0.197 621  &  0.012\%  \\
4  &  0.159 920  &  0.159 912  &  0.005\%  \\ \hline
\end{tabular}
\label{table}
\end{table}

The set of equations for the complete case is obtained by, first,
linearizing the system \eq{syst1} for infinitesimal adiabatic
perturbations and, second, by assuming the time and horizontal dependences
of the modes to be $\exp ({\rm i}\omega t)$ and $\cos (k_x x)$ or $\sin
(k_x x)$, respectively. We thus arrive at

\greq
\omega^2 \xi_x = k_x \Pi, \\ \\
\disp \omega^2 \xi_z = \dnz{\Pi} + N^2 \lp \xi_z + \frac{\Pi}{g}\rp,\\ \\
\disp \Pi = -c^2_s \lp -k_x \xi_x + \dnz{\xi_z} \rp - g \xi_z, \\ \\
\disp \xi_z = 0 \ltex{for} z=0,1.
\egreqn{full}
Here $\Pi = P'/\rho_0$ and $\rho'$ are the eulerian perturbations in
reduced pressure and density, respectively. This system has been obtained
under the classical Cowling approximation which assumes that perturbations
in the gravitational potential are negligible (Cowling 1941). We refer the
reader to the book of Unno et al.\ (1989) for a complete
derivation of this system in the case of a spherical geometry.

\begin{figure}
\centerline{\includegraphics[width=9cm]{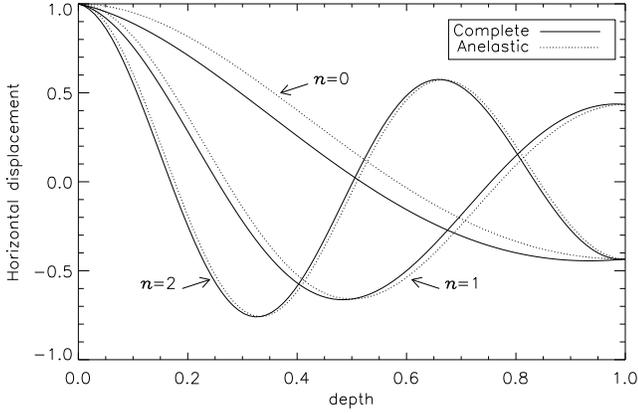}}

 \caption[]{Comparison of the complete (solid lines) and anelastic
 (dashed lines) normalized eigenfunctions $\xi_x$ for $\ell =1$ and
 $n=[0,1,2]$.}

\label{comp} 
\end{figure}

Lamb (1909, 1932) found analytical solutions of the system \eq{full} in
the case of an isothermal atmosphere. By eliminating $\xi_x$ and $\Pi$
in the system \eq{full}, we obtain a second-order differential equation
for $\xi_z$ only,

\greq
\disp \ddnz{\xi_z} + \frac{1}{H} \dnz{\xi_z} + \lc 
\frac{\omega^2}{c^2_s} - k_x^2 \lp 1 - \frac{N^2}{\omega^2} \rp \rc 
\xi_z = 0, \\ \\
\xi_z (0) = \xi_z (1) = 0.
\egreqn{lamb}
Comparison with Eq.~(\ref{eq_anel})
shows that the anelastic approximation applies provided that:

\begin{itemize}

\item the Lamb frequency term $\omega^2/c^2_s$ is negligible, implying that

\beq
\frac{\omega}{N} \ll k_x \frac{c_{\rm s}}{\omega},
\eeqn{condition}
which is a condition naturally fulfilled by high-order $g$-modes with
$\omega \ll N$.

\item the ratio $k_x c_{\rm s} / N$ is very large\footnote{
This last condition implies a clean separation between the acoustic
and gravity spectra such that a mixed ``gravito-acoustic'' mode is
not possible, i.e.\ the two spectra do not overlap (see Rieutord \&
Dintrans 2002).}

\end{itemize}

The solutions of the complete system \eq{lamb} can be deduced from the
anelastic ones as only the $\Delta$-term is changing, that is,

\beq
\Delta_{\hbox{\scriptsize compl}} = - \frac{1}{H^2} + 4 \lc 
\frac{\omega^2}{c^2_s} - k_x^2 \lp 1 - \frac{N^2}{\omega^2} \rp \rc,
\eeqn{delta_complete}
while the same quantization rule applies due to the identical boundary
conditions $\xi_z(0)=\xi_z(1)=0$, that is,

\beq
\Delta_{\hbox{\scriptsize compl}} = 4(n+1)^2\pi^2 = 4 k_z^2.
\eeqn{quant}
We thus obtain the following quadratic dispersion relation for the
complete eigenfrequencies

\beq
\omega^4 - c^2_s\lp k_x^2 + k_z^2 + \frac{1}{4H^2}\rp\omega^2
+ k_x^2 c^2_s N^2 = 0,
\eeqn{disper}
which is the same than that in Stein \& Leibacher (1974). In the same
way, the eigenvectors of the complete case can be deduced from
the anelastic ones and we find

\[ \left\{ \begin{array}{l}
\disp \xi^{\ell n}_x = \frac{C}{k_x} \exp(-z/2H) \lc k_z
\cos(k_z z) + \frac{1}{H}\lp\frac{1}{\gamma} - \frac{1}{2}\rp 
\sin (k_z z) \rc, \\ \\
\xi^{\ell n}_z = C \exp (-z/2H) \sin (k_z z).
\end{array}\right.
\]
Despite its different $\Delta$-term \eq{delta_complete}, we note that
in the complete case the
vertical displacement $\xi_z$ is the same as in the anelastic
one, which is simply a consequence of applying the same
quantization rule \eq{quant}.

We summarize in Table \eq{table} the eigenfrequencies for the complete case deduced
from \eq{disper} for
the first five $g$-modes of degree $\ell = 1$ 
and compare them with their anelastic counterparts given by the
formul\ae~\eq{wanel}. The agreement is quite remarkable since the
relative error done by the anelastic approximation is less than $1\%$
from the fundamental $g$-mode. Concerning the associated eigenvectors,
we showed that the vertical displacements $\xi_z$ are the same in both
cases so that only the horizontal ones differ.
However, Figure~\ref{comp} shows that the agreement on the normalized
$\xi_x$-eigenfunctions is also very good, except for the mode at $n=0$
for which the condition \eq{condition} is not well fulfilled.

In conclusion, the filtering of acoustic waves from the dynamics of the
isothermal atmosphere does almost not change the $g$-mode eigenfrequencies
$\omega_{\ell n}$ and eigenfunctions $\psi_{\ell n} = (\xi_x,\xi_z)^T$. The basis
built from these anelastic eigenfunctions can therefore be used with
good confidence to project out velocity fields from hydrodynamical simulations.

It should be pointed out that the application of the
anelastic subspace method will only give good results when the acoustic
and gravity spectra are well separated, that is, when the ratio $k_x c_s/N$
is large. This is the case for the chosen isothermal setup, where the
resulting anelastic eigenfrequencies/eigenvectors are very close to those
of the complete problem (see Table~\ref{table} and Fig.~\ref{comp}). However,
this ratio decreases with larger stratification, in which case one may be
forced to solve numerically the
oscillation equations \eq{full} of the complete problem instead of
those of the anelastic subset 
\eq{sysanel}, which does not however introduce any particular
difficulty.

\section{Damping rate of the modes using work integrals}
\label{A3}

We show in this appendix that it is possible to find the relation
governing the damping rate of a mode and the viscosity using the work
integrals formalism. We start from the anelastic equations written for
the velocity

\greq
\lambda \vu = - \na \Pi - N^2 \xi_z \ez + \nu \Delta \vu, \\ \\
\Div (\rho_0 \vu) = 0, \\ \\
u_z = 0 \ltex{for} z=0,1,
\egreq
where $\lambda = \alpha + i\omega$ has a non-zero real part due to the
viscous dissipation. We note that the viscous term is normally more
complicated as there may be a vertical variation of the dynamical
viscosity $\mu=\rho\nu$ [see the general form of this term in
Eq.~(\ref{syst1})]. However, we take here a simpler form as our aim is just
to formally prove that the damping rate $\alpha$ is linearly related to
the viscosity $\nu$.

We multiply the momentum equation by $\rho_0 \vu^*$ to obtain

\beq
\lambda \rho_0 \vu^2 = - \na (\rho \Pi \vu^*) - N^2 \rho_0 \xi_z u^*_z +
\nu \rho_0 \vu^* \cdot \Delta \vu,
\eeq
where we have used $\Div(\rho_0 \vu^*)=0$. As
in Appendix \ref{A1}, we finally integrate this last equation over the
volume and use the Green-Ostrogradsky theorem to obtain

\beq
\lambda \int_V \lp \rho_0 \vu^2 + N^2 \rho_0 \xi^2_z \rp {\rm d}V =
\nu \int_V \rho_0 \vu^* \cdot \Delta \vu {\rm d}V,
\eeq
where we also used $u^*_z = \lambda \xi^*_z$. As we only have real
integrals on both sides, we can deduce the following relation for the
real part $\alpha$

\beq
\alpha = \nu \frac{\int_V \rho_0 \vu^* \cdot \Delta \vu {\rm d}V}
{\int_V \lp \rho_0 \vu^2 + N^2 \rho_0 \xi^2_z \rp {\rm d}V}.
\eeq
The damping rate of a mode is therefore linearly related to the
viscosity, as found numerically in \eq{harm}.

A similar relation also applies in the complete case, provided one
includes the contribution of pressure fluctuations to the energy, that is,

\[
\alpha = \nu \frac{\int_V \rho_0 \vu^* \cdot \Delta \vu {\rm d}V}
{\int_V \lp \rho_0 \vu^2 + N^2 \rho_0 \xi^2_z + \frac{\Pi}{c^2_s} 
\rp {\rm d}V} \equiv \frac{\hbox{viscous dissipation}}{\hbox{total
energy}},
\]
where the total energy contains the kinetic ($\rho_0 u^2$), potential
($N^2 \rho_0 \xi^2_z$) and internal ($\rho_0 \Pi/c^2_s$) contributions.
Such relations are useful from a numerical point of view as they
allow to check the accuracy of the computed eigenvalues from their
associated eigenvectors.

\section{Parseval-Bessel relation in the anelastic subspace}
\label{A4}

In this appendix, we will show that it is possible to relate the total
kinetic energy embedded in the box to the one calculated in the anelastic
subspace. To do that, we need a similar relation than the Parseval-Bessel
one which states that the total energy is conserved when one works in
Fourier space, i.e.\

\beq
\int_V f^2(x,y,z) \,{\rm d} V = \int_k |\hat{f}|^2 {\rm d}^3k,
\eeq
where $\hat{f}$ denotes the Fourier transform of $f$.

In our case, we only perform a 1-D Fourier transform of the velocity
field in the horizontal direction (corresponding to a wavenumber
$k_x = (2\pi/L_x)\ell$), so that the Parseval-Bessel relation for the
total kinetic energy at a given time is

\beq
\int_V \rho_0 \vi^2(x,z) \,{\rm d}x \,{\rm d}z = \int_z
\sum^{+\infty}_{\ell} 
|\hat{\vi}_\ell|^2 \rho_0 \,{\rm d}z,
\eeqn{bessel1}
where $\hat{\vi}_\ell$ denotes the velocity field at a wavenumber $k_x$.
We now introduce the anelastic subspace by using the relation which
links the velocity field in Fourier space to the one in the anelastic
space as

\beq
\hat{\vi}_\ell (z) = \sum^{+\infty}_{n} \langle \vpsi_{\ell
n},\hat{\vi}_\ell
\rangle \vpsi_{\ell n} (z) + {\rm rest}.
\eeq
The $z$-integral in Eq.~\eq{bessel1} can then be calculated in
the anelastic subspace as

\beq
\begin{array}{ll}
\disp \int_z |\hat{\vi}_\ell|^2 \rho_0 \,{\rm d}z = 
\int_z \hat{\vi}^\dagger_\ell \hat{\vi}_\ell \rho_0 \,{\rm dz}, \\ \\
\disp \hspace{0.5cm} = \sum^{+\infty}_{n} \langle \vpsi_{\ell
n},\hat{\vi}_\ell
\rangle \int_z \hat{\vi}^\dagger_\ell \vpsi_{\ell n} (z) \rho_0 \,{\rm d}z 
+ {\rm rest}, \\ \\
\disp \hspace{0.5cm} = \sum^{+\infty}_{n} \langle \vpsi_{\ell
n},\hat{\vi}_\ell
\rangle \langle \hat{\vi}_\ell,\vpsi_{\ell n}\rangle  + {\rm rest}, \\ \\
\disp \hspace{0.5cm} = \sum^{+\infty}_{n} |c_{\ell n}|^2 + {\rm rest}.
\end{array}
\eeq
This last equation means that the total kinetic energy due to $g$-modes
at a given $k_x$ and time $t$ is simply given by the sum over the radial
orders $n$ of the squares of the mode amplitudes. By taking into account
the $k$-integral, we finally obtain the Parseval-Bessel relation written
in the anelastic subspace as

\beq
\int_V \rho_0 \vi^2(x,z) \,{\rm d}x \,{\rm d}z = \sum^{+\infty}_{\ell,n}
|c_{\ell n}|^2 + {\rm rest}.
\eeq

\end{document}